\documentclass[%
 reprint,
 amsmath,amssymb,
 aps,
]{revtex4-2}

\usepackage{graphicx}% Include figure files
\usepackage{dcolumn}% Align table columns on decimal point
\usepackage{bm}% bold math
\begin{document}

\preprint{APS/123-QED}

\title{Symmetry breaking Paradigm In Typical Laminar-Turbulence Transition System}% Force line breaks with \\

\author{Chun Huang}
 \email{chunhuang@mails.ccnu.edu.cn}
 \affiliation{%
 Physics Department, Central China Normal University,152 Luoyu Road, Wuhan, Hubei 430079, China
}%
\affiliation{%
 Physics Department, Washington University in St. Louis,  1 Brookings Dr, St. Louis, MO 63130, US
}%
\author{Yuchen Jiang}%
\affiliation{%
 Physics Department, Central China Normal University,152 Luoyu Road, Wuhan, Hubei 430079, China
}%

\date{\today}% It is always \today, today,

\begin{abstract}
A stationary cylindrical vessel containing a rotating plate near the bottle surface is partially filled with liquid. With the bottom rotating, the shape of the liquid surface would become polygon-like. This polygon vortex phenomenon is an ideal system to demonstrate the Laminar-Turbulent transition process. Within the framework of equilibrium statistical mechanics, a profound comparison with Landau's phase transition theory was applied in the symmetry-breaking aspect to derive the evolution equation of this system phenomenologically. A comparison between theoretical prediction and experimental data is carried out. We concluded a considerably highly matched result, while some exceptions are viewed as the natural result that the experiment breaks through the up-limit of using equilibrium mechanics as an effective theory, namely breaking through the Arnold Tongue. Some extremely complex Non-equilibrium approaches were desired to solve this problem thoroughly in the future. So our method could be viewed as a linear approximation of this theoretical framework.
\end{abstract}

\maketitle

%\tableofcontents

\section{Introduction}
The polygonal vortex phenomenon refers to the phenomenon that the free surface of a partially filled cylindrical container with a rotating plane near the bottom appears a polygonal shape with the rotation of the plane. This phenomenon could be divided into dry polygon and wet polygon by whether there is water or not in the core region upon the bottom plane. With the varying initial height of the liquid and rotating frequency, these two different forms of polygon vortex could be seen separately.

The polygon-like free surface formation was first reported by Vatistas(1990) \cite{VatistasFirstExperiment}.In that paper, the researchers reported periodic sloshing phenomena with the polygon formation. After this qualitative experiment, Some more detailed and quantitative experiments also were carried out with other collaborators \cite{Vatistas2Experiment}\cite{Vatistas3Experiment}\cite{ExperimentalResearch}. Furthermore, some experimental research even broadened the boundary of this phenomenon in extremely huge scale and non-viscosity system\cite{hugescale1}\cite{hugescale2}\cite{hugescale3}.

On the other hand, some theoretical research also is carried out by other groups. One of these works is named the Point Vortex model, which inherits the work of Lord Kelvin \cite{PointVortexPromotion}\cite{VatistasLinearStability}. However, some researchers doubt the strong assumption and limitations of this model in the further development\cite{VatistasViscosityNOSENSE}. Another candidate is the Wave Resonance model which is regarded as the most successful model in this field nowadays, first developed by Tophøj(2013)\cite{Topj:firstModelDevelop}. After that, J Mougel went further and solved the sloshing and switching problem during the evolution process within the framework of this model\cite{SloshingProblemMogoul}. Then, they promote this model even further to solve the global stability problem in common parameter space.\cite{GlobalStabilityMogoulBohr}

However, the symmetry-breaking idea of this system reminds us that the introduction of a standard phase transition paradigm in equilibrium statistical mechanics could be promising. This viewpoint would be almost purely non-hydrodynamical, macroscopic, and phenomenological while also catching the physical significance as this ideal Laminar-Turbulence transition process generally is a non-equilibrium mechanical problem.

In this writing, we will first derive a profound promotion of Landau's phase transition theory to get the evolution equation of handset order parameter in a transcendental way in Section.2, then we will utilize the most successful Wave Resonance framework to recognize a physically solid base for the handset order parameter in Section.3. Some fascinating outlooks about this topic with non-equilibrium statistical viewpoints are put forward in Section.4.

\section{Transcendental Promotion}
During the evolution history of this Polygon Vortex system, two different stages would be seen. One of them is the axisymmetric stage. In this stage, the general shape of the free surface is symmetric concerning the central axis of the cylindrical vessel. While within the spin-up process, this geometric symmetry would be broken, and get into the non-axisymmetric stage, in which, the polygon vortex phenomenon would appear so that we conclude that the system gets through symmetry breaking process phenomenologically, very similar to the phase transition in equilibrium statistical mechanics, and we can similarly name the different stages as 
\textit{Axisymmetric phases} and\textit{ Non-axisymmetric phases}.
Then the evolution history of this system would be simplified to a phase transition process, and the dominant factor is the symmetry of the free surface. This symmetries broken phenomenon characterizes the occurrence of phase transition.

The free energy $ G $ in Landau’s theory was replaced by the total gravitational potential $ E $ in this system. Inspired by the phenomenon that every time the pattern transfer from one to the other, the drop of the liquid height would be seen in the container\cite{Newton}, we consider that at that point must exist a local minimum in gravitational potential, which has a similar behavior compared with the free energy in the phase transition theory.

Then a handset order parameter named \textit{Maxima frequency}
, was defined as in the polar coordinates the oscillation frequency of the intersection between the polygon pattern and the given polar angle radius. To characterize the different rotation orientations, we defined it as the absolute value of a vector that has the same orientation with the angular momentum. The mathematical form of this parameter is $\boldsymbol{M}=f_{M} \hat{L}$, $\hat{L}$ is the unit vector along the angular momentum direction.
More specifically, we can rigorously define this order parameter as
\begin{equation}
\boldsymbol{M}=\frac{n \boldsymbol{\omega}}{4 \pi}
\label{1}
\end{equation}
which is also concise in form. where n is the number of the sides, $\boldsymbol{\omega}$ is the angular velocity.
This definition satisfies the sufficient conditions of being an order parameter, which means it remains 0 when the system is still highly symmetric while the number jumps to a non-zero value after a symmetry-breaking process happened.
Besides, frequency $f_M$ is proportional to the number of the sides $ n $, which is a theoretical formulation for the experimental results of frequency locking\cite{frequencylocking}. The physical origin analysis would be shown in the next section, by far, we just view it as a handset pragmatic parameter.

Following is the symmetry analysis part. First expanding the gravitational potential with the Taylor series (near the $\boldsymbol{M}=0$)
$E=\sum_{j=0}^{\infty} L_{j} \cdot(\boldsymbol{M})^{j}$,
in which $ L_{j} $ are the expanding coefficients. Expanding gravitational potential near the $\boldsymbol{M} = 0$ points means that this operation is done in the neighboring region of the symmetry breaking point.

\begin{figure*}
	\centering
	\includegraphics[scale=0.4]{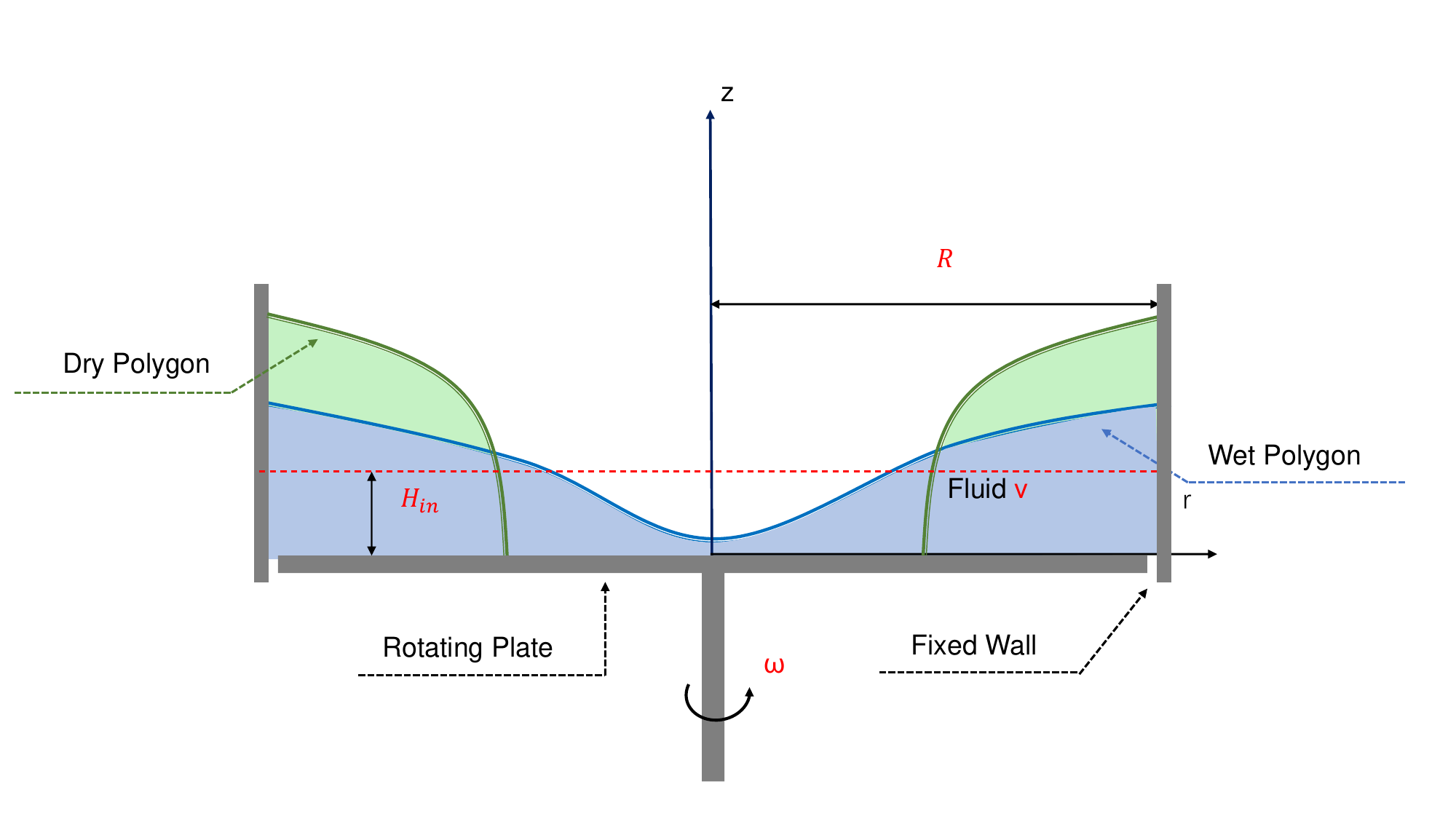}
	\caption{\textsl{Sketch of the system set with parameters (Cross-sectional view), featuring the bottom plate and side wall, with different fluid patterns demonstrates both dry polygon and wet polygon
	}}
	\label{fig.1}
\end{figure*}

Due to the rotation symmetry of the system and the formula must be scalar, all the odd terms should be canceled out. So the Taylor series reduced to
\begin{equation}
E=\sum_{j=0}^{\infty} L_{2 j} \cdot(\boldsymbol{M})^{2 j}
\end{equation}
When $f \leq f_{c}$, $E=L_{0}(f)$, where $ f $ is rotation frequency, $ f_{c} $ is critical phase transition frequency. Calculating the $L^{wet}_{0}$ term out of the Euler equation for the wet polygon. Then for some super shallow situations, it could still be axisymmetric even after the dry area appears. Thus we derive the $L^{dry}_{0}$ term for dry polygon by volume conservation and Navier-Stokes equation(see Figure \ref{fig.1}). When we take the derivation procedure to get the minimal, the messy $L_{0}$  can always be omitted, resulting in the irrelevance of the final results. (detailed derivation about  $L_{0}$ would be found in Supplementary material.1)

After that, we could omit all the items of $ E $ above the fourth order as a low order approximation,
\begin{equation}
E=L_{0}(f)+Cu \cdot \nu\left(f_{c}-f\right)^{-1}\boldsymbol{M}^{2}+H_{in} R\left(f_{c}-f\right)^{-2}\boldsymbol{M}^{4}
\end{equation}
in this formula, $f_c$ is the critical frequency of phase transition, $\nu$ is the kinematic viscosity of the fluid, $H_{in}$ is the initial height of the liquid in the container, $Cu$ is a dimensionless quantity that is originated from some solid physical considerations:
\begin{enumerate}
	\item The coefficient of the quadratic term illustrates the dynamical constraints for the system and the dimensionless number $ Cu $ combine the effect of Coriolis force, gravitational force, centrifugal force, viscous force, and momentum force.
	\begin{equation}
	Cu=\frac{(\textit{Coriolis force})^{3}(\textit{viscous force})}{(\textit{gravitational force})^{2}(\textit{momentum force})^{2}}
	\label{4}
	\end{equation}
	\item The coefficient of the fourth order term reflects the spatial restriction of the evolution of the polygon. For both $H_{in}$ and R is the main spatial scale dominators of this fluid system.
	\item  The overall tool to determine the exponent is dimensional analysis.
\end{enumerate}
After getting the expression of $ E $ for the order parameter $\boldsymbol{M}$. The basic picture would become clear: with the increase of rotation frequency, the local minima, which initially is located at the $\boldsymbol{M}=0$, would split to other points, as the physical meaning of local minima is referred to as the physically realizable phase of the system. So accompanied by the splitting behavior, the phase transition happens. To find the local minima and get the dynamic information of them to rotation frequency, with the extreme condition $\partial E/\partial \boldsymbol{M} =0$, we could get
\begin{equation}
\boldsymbol{M}=0 \qquad
\boldsymbol{M}=\pm{\left[  -\frac{L_{2}(f)}{2 L_{4}(f)}\right]  }^{\frac{1}{2}}\hat{L}
\end{equation}
These two solutions demonstrate to us the accessible local extremes. Then, with $\partial^{2} E/\partial \boldsymbol{M}^{2}>0$, we can gain the information that in which frequency interval, $ E $ would be the local minima.
\begin{equation}
\boldsymbol{M}(f)=\left\{\begin{array}{ll}
0 & \left(f \textless f_{c}\right) \\
\left[  {\frac{1}{2} \frac{C u \cdot \nu}{H_{in} R}\left(f -f_{c}\right)}\right]  ^{\frac{1}{2}}\hat{L} & \left(f \geq f_{c}\right)
\end{array}\right.
\label{6}
\end{equation}
So that means when the frequency is lower than the critical phase transition value, the order parameter would remain zero. 
When the frequency is larger than the critical value, the transition happened. The value of $\boldsymbol{M}$ would be determined by the next formula, and as the $\boldsymbol{M} \propto n$ in (\ref{1}), the sides number of the polygon would be uniquely specified.

Taking (\ref{1}) into (\ref{6}), the formulation of the sides number of the polygon n could be derived concerning the rotation frequency of the bottom plane.
\begin{equation}
n=\frac{1}{2} T_{p o} \left[ {\frac{C u \cdot \nu}{2 H_{in} R}\left(f-f_{c}\right)}\right]  ^{\frac{1}{2}}
\end{equation}
where $T_{po}$ is the rotating period of the polygon-like free surface, this formula would reflect how the number of the sides would evolute as the frequency increase. The dimensionless number Cu, initial height H, critical frequency, and container radius R are the vital quantities that determine the evolution of the polygon vortex.

\begin{figure}[h]
	\centering
	\includegraphics[scale=0.4]{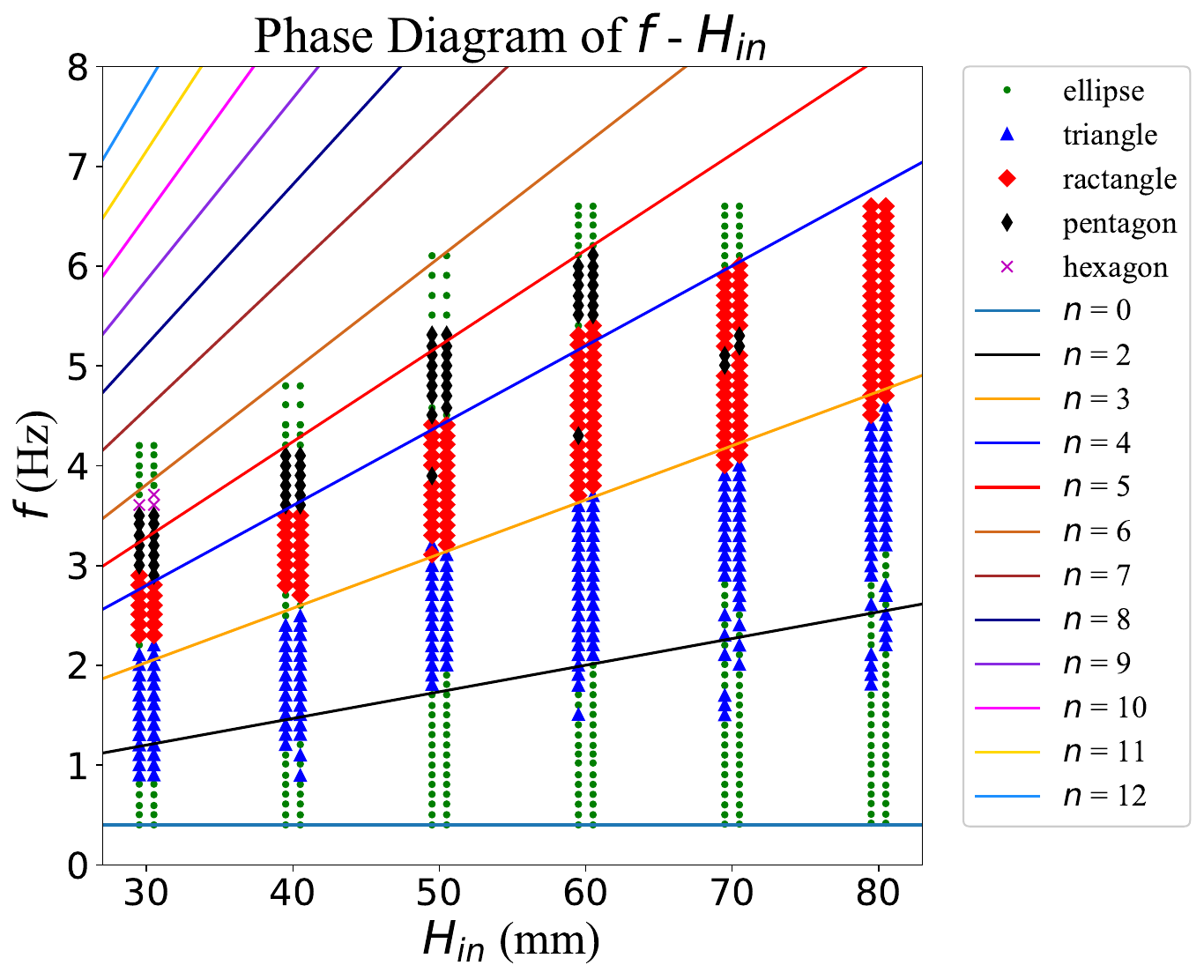}
	\caption{\textsl{Thanks to the data of DOI$\colon$10.1017/jfm.2014.568. In this phase diagram, we label experimental data as different point patterns. And the different sides numbers could be predicted by our model, labeled by various colored lines, where n is the number of the sides}}
	\label{fig.2}
\end{figure}

Using the experimental data of  B.Bach's group\cite{Newton}, we performed the comparison between the experiment and the prediction of our model in Figure \ref{fig.2}.
Therefore we conclude that with only one adjustable parameter in the formula and all the result that comes from the general symmetry feature analysis of this system, the performance of our model is incredibly excellent. What's more, as the initial value could be uniquely determined by (\ref{4}), we find out the following numerical value of $ Cu $ could be written by a concise form $ Cu = kn $  after analyzing the experimental data, which may also be true under different initial conditions with corresponding constant $ k $. For we don't know exactly why it is so effectual, especially with a handset order parameter, some further analysis from the fluid dynamics origin aspect would be beneficial and necessary.

\section{Physical Origin} 
In this section, the solid base of our physical idea would be explored, which originated from the Wave Resonance theory, in which we recognized a vital physical quantity named dispersion relation that has the same structure of satisfying the conditions of being an order parameter. Furthermore, their similar polygonal system evolution behaviors reveal the same physical significance between them. So we assume it as the profound origin of this model.

To analyze the free surface generated wave motion, a velocity potential perturbations would be added by $\vec{v}_{p}=\nabla \Phi $ with the method inheriting from Tophøj(2013)\cite{tophoj2013rotating}. This potential would satisfy the Laplace equation.

With the free surface boundary condition in cylindrical coordinates,  with its definition, the flow velocity of azimuthal direction can be experessed by $U(r)=(\Gamma/2 \pi r) \hat{\theta}$, where $\Gamma$ is the circulation.

By solving the Laplace equation, we got two different solutions and we identified the gravity waves family and centrifugal waves family separately.
\begin{equation}
\begin{aligned}
\Phi_{c} &=\left[K_{1}\left(\frac{r}{R}\right)^{m}+K_{2}\left(\frac{r}{R}\right)^{-m}\right] e^{i(m \theta-\omega t)}\\
&\Rightarrow\textit{gravity waves family}\\ 
\Phi_{g} &=\left(K_{3} e^{m z / R}+K_{4} e^{-m z / R}\right) e^{i(m \theta-\omega t)}\\
&\Rightarrow\textit{centrifugal waves family}
\end{aligned}
\end{equation}
Where $ m $ is the wave number the same as $ n $ in our model(another geometrical parameter could be found in Figure \ref{fig.1}). Then with the free surface boundary condition, we could constrain these solutions to a governing equation of the waves families
\begin{equation}
\begin{aligned}
&\left[\left(\omega - m \frac{V(R_{d})}{R_{d}}\right)^{2}-g_{c}m F/R_{d}\right]\times\\
&\left[\left(\omega - m \frac{V(R)}{R}\right)^{2}-g m F/R\right]
=\frac{m^2 g g_c}{R_{d} R}(F^{2}-1)
\end{aligned}
\label{9}
\end{equation}
in which $g_c$ is the centrifugal acceleration $g_{c}=\Gamma^2/(4\pi^2 R_{d}^3)$, where F is 
\begin{equation}
F = \frac{1+e^{-2m \zeta/R}\left(\frac{R_{d}}{R}\right)^{2m}}{1-e^{-2m \zeta/R}\left(\frac{R_{d}}{R}\right)^{2m}}
\end{equation}
As was mentioned in Tophøj(2013)[18], for $(F^{2}-1)$ is a comparatively small quantity in a large m limit, we recognize the dispersion relations for the two families of waves that
\begin{equation}
\begin{aligned}
D_{g}(\omega)&=\left[\left(\omega - m \frac{V(R)}{R}\right)^{2}-g m F/R\right]\\
&\Rightarrow\textit{Gravity waves dispersion relations} \\
D_{c}(\omega)&=\left[\left(\omega - m \frac{V(R_{d})}{R_{d}}\right)^{2}-g_{c}m F/R_{d}\right]\\
&\Rightarrow\textit{Centrifugal waves dispersion relation}&
\end{aligned}
\end{equation}
With a more careful identification, we can clarify the characteristics of the dispersion relations.
\begin{enumerate}
	\item They are separately related to different wave families and the product of these quantities could be the dominator of this system. The remaining procedure is to solve the equation (\ref{9}).
	\item When the free surface remains axisymmetric, their value all would be zero, while symmetry is broken, they will get some non-zero values. The sides number of the polygon vortex is directly reflected in these quantities with the wave number m.
	\item The dimensionality of these dispersion relations was $\left[T\right]^{-2}$, which is the square of the dimensionality of our handset order parameter.
\end{enumerate}
For the 1-st and 2-nd features, we can conclude that the product of these dispersion relations could be seen as an order parameter, and with the 3-rd characteristic, we can conclude that the dispersion relations are power-correlated with our order parameter. So the main governing equation from purely Hydrodynamics aspect could be written as this form
\begin{equation}
D_{s}(\omega)\equiv D_{g}(\omega)D_{c}(\omega)=\frac{m^2 g g_c}{R_{d} R}(F^{2}-1)
\end{equation}
In which, we defined a dominate order parameter of this whole system with the notation $D_{s}(\omega)$. Compared with our handset order parameter $\boldsymbol{M}$ by the way of dimensional analysis, we can find the relation:
\begin{equation}
\left[D_{s}(\omega)\right]^{\frac{1}{4}} \propto m^{\frac{1}{2}}
\end{equation}
Then with the relation $ C u = kn $ mentioned before, we can also find the relation in $\boldsymbol{M}$ 
\begin{equation}
\lvert\boldsymbol{M} \rvert\propto n^{\frac{1}{2}}
\end{equation}

Thus, it's obvious that the evolution of $\left[D_{s}(\omega)\right]^{\frac{1}{4}}$ and $ \vec{M} $ is subject to changes in $ m^{\frac{1}{2} }$ and $ n^{\frac{1}{2} }$, respectively. While the wave number $ m $ is the same as the number of the sides $ n $ in our model, we can conclude that the Wave Resonance theory of Tophøj can be simplified as the work of seeking the main order parameter in this system. 
So we just determine the dominant order parameter in two different ways. Compared with these two order parameters, the handset one even is almost purely based on our physical intuition of ourselves, while still owning its irreplaceable advantages. For instance, as a vector, the rotating orientation could be reflected but on the other side, this direction information wouldn't influent the final result which illustrates the symmetry of the system for the direction of that rotation. Furthermore, as the hydrodynamics derivation is based on the potential flow hypothesis and Bernoulli principle, the physical assumptions would be regarded as extremely strong confinement to the power of this order parameter from the Wave Resonance framework.

\section{Conclusion}
%while as a typical macroscopic symmetry breaking system, the purely phenomenological model is absent unexpectedly. 
As all the previous research in this field are hydrodynamical-focusing, inspired by Landau's second order phase transition theory, one of the motivations of this work is to fill the blank of the modeling which is constructed systematically based on the symmetry transformation behaviors.
% an effective comparison was gained. %两句动机可以简化一下
However, with the advance in the investigation, we generally realized the intrinsic focus of this model we built as it is following the paradigm of equilibrium statistical mechanics. Since this system could be identified as the classical non-equilibrium states in \textit{polygon vortex} phase space, %实际上是相空间
the model we derived always could be seen as using the equilibrium phase transition paradigm to investigate the non-equilibrium system. %我们的模型是用平衡相变方法来研究非平衡系统
Some clues imply this belief clearly: multi-stability \cite{multistability} as the time-inversion \& phase-space-inversion symmetry violation due to the small perturbations during the evolution process; The stable states, Kelvin equilibrium, could be seen as the Limit cycle's attraction points during the evolution. The phase point in the phase space will follow the phase track absorbed in the limit cycle and falls into the attraction points. Then with increasing the perturbation level, this phase will be transformed into other stable states. %引用一下相关书籍
And by definition, the diversity of the non-equilibrium system behaviors could be highly determined by the initial setting of the parameter space volume. When the outer parameter reached some specific value, the Arnold tongue would be broken through, and the quasi-periodical behaviors that the system demonstrated before will be violated to the undeterminable chaos. The applicable scale of our model is limited to the periodical and quasi-periodical phase space. %最后一句说明我们的理论在哪个层次
In the chaotic system, multiple order parameters could be helpful as Haken suggested in 1980s\cite{SynergeticsHaken}. Or a more valid order parameter intrinsic to this system and even on Laminar-Turbulence transition would be called as complexity science suggested. These fancy concepts already exist in this some intelligent experimentists' data analysis\cite{FirstUSEofNONlinearmethod}, referring to the same structure as we mentioned. Some also attempted to apply them in much less complicated system modelling\cite{last1}\cite{last2}\cite{last3}.
%我们先说相关的概念的研究在一些相似的简单系统中已经被证明是成功的了，但是由于系统过于复杂，目前没有人用这些概念在这个多边形涡系统中，所以我们要建立起这个模型来窥探这个模型
But to chaos and complexity, no research ever be implemented in this direction. The model in this writing could inspire the following research on the ultimate model. Along the direction of our model, a more powerful and elegant unified theory could be completed.
\subsection*{Data Availablity}
The dataset is available in B. Bach, E. C. Linnartz, Malene Louise Hovgaard Vested, Anders Peter Andersen, and Tomas Bohr. From newtons bucket to rotating polygons: experiments on surface instabilities in swirling flows. Journal of Fluid Mechanics, 759:386–403, 2014. DOI$\colon$10.1017/jfm.2014.568.
\subsection*{Conflict of Interest}
The authors have no conflicts to disclose. We declare that we do not have any commercial or associative interest that represents a conflict of interest in connection with the work submitted.

\bibliography{citation.bib}% Produces the bibliography via BibTeX. 这个包含引用文献的.bib文件是需要自己根据所引用的文章用文献管理工具或者自己手工生成的。
\end{document}